\documentclass[conference]{IEEEtran}
\IEEEoverridecommandlockouts
\usepackage{cite}
\usepackage{amsmath,amssymb,amsfonts}
\usepackage{graphicx}
\usepackage{epsfig}
\usepackage{epstopdf}
\usepackage{textcomp}
\usepackage{xcolor}
\usepackage{float}
\usepackage{soul}
\usepackage{authblk}
\usepackage{algpseudocode}
\usepackage{algorithm}
\usepackage{subcaption}
\usepackage{multirow}
\usepackage{array}
\usepackage{booktabs}

\def\BibTeX{{\rm B\kern-.05em{\sc i\kern-.025em b}\kern-.08em
    T\kern-.1667em\lower.7ex\hbox{E}\kern-.125emX}}
\begin{document}

\title{MCS Adaptation within the Cellular V2X Sidelink\\}

\author[*]{Andres Burbano-Abril}
\author[$\dag$]{Brian McCarthy}
\author[$\ddag$]{Miguel Lopez-Guerrero}
\author[*]{Victor Rangel}
\author[$\dag$]{Aisling O'Driscoll}

\affil[*]{\textit{School of Engineering}, \textit{National Autonomous University of Mexico}, Mexico City, Mexico}
\affil[$\dag$]{{\textit{School of Computer Science and Information Technology}, \textit{University College Cork}, Cork, Ireland}}
\affil[$\ddag$]{\textit{Department of Electrical Engineering}, \textit{Metropolitan Autonomous University-Iztapalapa}, Mexico City, Mexico}
\affil[*]{bburbano@comunidad.unam.mx, victor@fi-b.unam.mx}
\affil[$\dag$]{\{b.mccarthy, a.odriscoll\}@cs.ucc.ie}
\affil[$\ddag$]{milo@xanum.uam.mx}

\maketitle
\begin{abstract}

Adaptation of the Modulation and Coding Scheme (MCS) within the Cellular Vehicle-To-Everything (C-V2X) sidelink has the potential for a wide range of applications including congestion control, support of variable packet sizes, density aware rate adaptation and improved support of unicast transmissions. However, the practical implementation of MCS adaptation presents a wide range of implications for the C-V2X radio resources, bandwidth, computation of power levels and operation of the scheduling mechanism. This paper presents the first study that provides a detailed analysis and an implemented model highlighting the implications of MCS adaptation on the operation of the Sensing-Based Semi-Persistent Scheduling (SB-SPS) mechanism within the C-V2X sidelink. To showcase the use of MCS adaptation for a particular purpose, a detailed analysis of its performance for distributed congestion control is undertaken, while considering different vehicular densities. The results indicate that MCS adaptation can be useful to reduce channel congestion by decreasing resource occupation but may not improve the overall packet delivery rate unless subchannel occupation is reduced. Finally, this study provides the foundation for other applications of MCS adaptation within the C-V2X sidelink.
\end{abstract}

\vspace{-9pt}
\section{Introduction}
\label{section-introduction}
\vspace{-3pt}
The advent of vehicular communication technology is set to revolutionize the transport \& mobility sector, enabling improved safety and traffic efficiency. Beyond this, the increase in vehicular autonomy supported by enhanced sensing capability is poised to deliver new advanced services including cooperative sensing and maneuvers, many of which will utilize direct sidelink connectivity. Over the last decade, vehicular communications standards were based exclusively on the IEEE 802.11p technology standards \cite{4526014}. However, more recently the Third Generation Partnership Project (3GPP) has rapidly specified cellular standards to address the challenges of vehicular communications, starting with Release 14 \cite{3gpp-TS-36-885} (also known as LTE-V or C-V2X) and followed by Release 15 \cite{3gpp-TR-21-915} and Release 16 \cite{3gpp-TR-21-916}, commonly referred to a NR-V2X. 

Communication between vehicles and their environment can be supported either by the cellular network infrastructure (C-V2X Mode 3 or NR-V2X Mode 1) or by allowing vehicles to communicate directly (C-V2X Mode 4 or NR-V2X Mode 2) in an ad-hoc fashion. In the vehicular sidelink, vehicles can autonomously allocate and manage a fixed amount of resources for predefined periods of time, using a scheduling mechanism known as Sensing Based- Semi Persistent Scheduling (SB-SPS). While resources are reserved in a semi-persistent basis, vehicles can still modify parameters like the transmission power or the MCS. Therefore, vehicles can adapt these parameters for different purposes such as congestion control \cite{etsi-congestionControl}, the transmission of packets of variable size \cite{8676227}, among others.\vspace{-1pt}

However, while adapting the transmission power during ongoing transmissions in the C-V2X sidelink is straightforward because it does not modify the resource occupation, adapting the MCS is a more challenging task. This is because different MCS configurations modify the number of transmission resources and consequently the transmission bandwidth. This has further implications in the computation of the power levels in the channel and in the operation of SB-SPS, which relies on specific power metrics in order to select and reserve the required resources. As yet, a detailed analysis of the implications of adapting the MCS within the vehicular cellular sidelink have not been addressed in literature.

This paper addresses this challenge by presenting a comprehensive analysis of the implications of MCS adaptation on the operation of the C-V2X sidelink. Specifically, it analyzes the impact of MCS adaptation in the resource organization of C-V2X, as well as in the computation of the power levels in the channel and their implications in the operation of SB-SPS. It further analyzes and accurately models the performance implications of adapting the MCS for distributed congestion control, which to the best of the authors' knowledge has not been explored before. Finally, it provides the foundations for other studies that want to utilize the implementation of MCS adaptation for other purposes.

The remainder of this paper is organized as follows. Section \ref{section-literature} discusses related literature and Section \ref{section-background} describes the background and operation of C-V2X Mode 4. Section \ref{section-modeling} describes the implications of MCS adaptation on the operation of SB-SPS and discusses the intricacies of modeling this within C-V2X. Section \ref{section-performance} presents the performance of MCS adaptation for distributed vehicular congestion control. Finally, Section \ref{section-conclusions} provides some conclusions.

\vspace{-2pt}
\section{Related Literature}
\label{section-literature}
\vspace{-3pt}
Adapting the MCS in real-time has mainly been proposed as a mechanism for distributed congestion control in C-V2X, as flagged by the 3GPP and the related literature. To a lesser extent, it has also been proposed as a potential solution for transmitting packets of variable size.

However, studies in the area have either investigated the impact of different static MCS configurations which is not in compliance with ETSI standards \cite{etsi-congestionControl}, or have focused on discussing the potential of MCS adaptation without analyzing its operation and performance. For instance, the authors in \cite{8080373} and \cite{8795500} briefly discuss the use of MCS adaptation for congestion control in C-V2X Mode 4 and analyze how increasing the MCS can potentially reduce congestion by reducing the number of required resources at the expense of a decrease in performance. However, these works do not analyze the implications of implementing this mechanism on the operation of C-V2X SB-SPS, which is crucial for it's performance.

The study in \cite{9048738} presents a detailed analysis of different congestion control techniques in C-V2X and shows the impact of different MCS configurations on the performance of C-V2X when vehicular congestion is present. However, the study assumes that MCS configurations remain fixed during the simulations and does not analyze the operation of MCS adaptation for congestion control following the ETSI specifications \cite{etsi-congestionControl}. Such standards require that MCS adaptation operate dynamically during ongoing transmissions and as function of the congestion conditions in the channel.

Other studies, such as \cite{8676227} and \cite{9133075}, discuss the applicability of MCS adaptation as a solution for transmitting vehicular packets of variable size using C-V2X. These works provide an analysis of the implications of MCS adaptation and highlight the relevance of the technique for this particular application. However, they do not model the operation of MCS adaptation, which is critical to assess the performance of this technique when the size of the packets is variable.

MCS adaptation can have multiple applications in C-V2X beyond those already discussed in the related C-V2X literature. This is demonstrated by literature utilizing the wireless 802.11p based vehicular standard. For instance, \cite{6823631} and \cite{7480406} show how MCS adaptation can be used to improve the performance of broadcast transmissions by adapting the MCS to variable channel conditions. Similarly the authors in \cite{5487432} highlight the relevance of MCS adaptation on the performance of vehicular unicast transmissions, which is a feature of NR-V2X under the 5G Release 16 specification \cite{3gpp-TR-21-916}. 

\section{C-V2X Overview}
\label{section-background}

The C-V2X standard builds upon the sidelink PC5 interface, originally designed for Device-To-Device (D2D) communications in LTE. but introduces new physical (PHY) and medium access control (MAC) layers to guarantee an adequate performance of V2X communications vehicular scenarios. 

\subsection{C-V2X Physical Layer}
\label{subsection-phy}

The physical layer of C-V2X implements Single-Carrier Frequency Division Multiple Access (SC-FDMA), where resources that are organized in a grid structure can be either allocated by the eNodeB or autonomously selected by the vehicles. 

In the time domain, resources are organized in frames of 10 ms, which are further divided into subframes of 1 ms. Each subframe contains up to 15 SC-FDMA symbols, where 9 symbols carry data and 6 symbols are reserved for the transmission of control information. In the frequency domain, the total bandwidth can be configured to be either 10 MHz or 20 MHz. The total bandwidth is further divided into subchannels consisting of a group of Resource Blocks (RBs), where each RB comprises 12 subcarriers of 15 kHz (180 kHz total). Under this structure both the number of subchannels and their size are configurable but depend on the available bandwidth.

The Physical Sidelink Control Channel (PSCCH) occupies 2 RBs and is used to transmit the Sidelink Control Information (SCI), which contains the information required by the receiver to decode each transmission. The Physical Sidelink Shared Channel (PSSCH) transmits the data in Transport Blocks (TBs) and its size dependant on packet size and the MCS. RBs can be allocated contiguously (adjacent scheme) or they can be allocated in separate pools (non-adjacent mode).

\subsection{C-V2X MAC Layer}
\label{subsection-mac}

Resources can be either allocated by the cellular network infrastructure (Mode 3) or autonomously selected by the vehicles (Mode 4). In Mode 4, vehicles first sense the channel and then select one or multiple subchannels for a given number of consecutive transmissions using the SB-SPS mechanism.

Every time a packet arrives from the upper layers and a reservation is not in place, the SB-SPS mechanism generates a new resource grant. This grant contains information regarding the number of subchannels to be reserved, the duration of the reservation in terms of the number of consecutive transmissions and the period between them. The period between transmissions is set in the Resource Reselection Interval (RRI) parameter and is fixed due to the periodic behavior of SB-SPS. In turn, the number of transmissions is chosen at random within a range of values that depend on the RRI and is set in the Resource Reselection Counter (RRC) parameter.

\begin{figure}[htbp]
\centering
\includegraphics[scale=0.38]{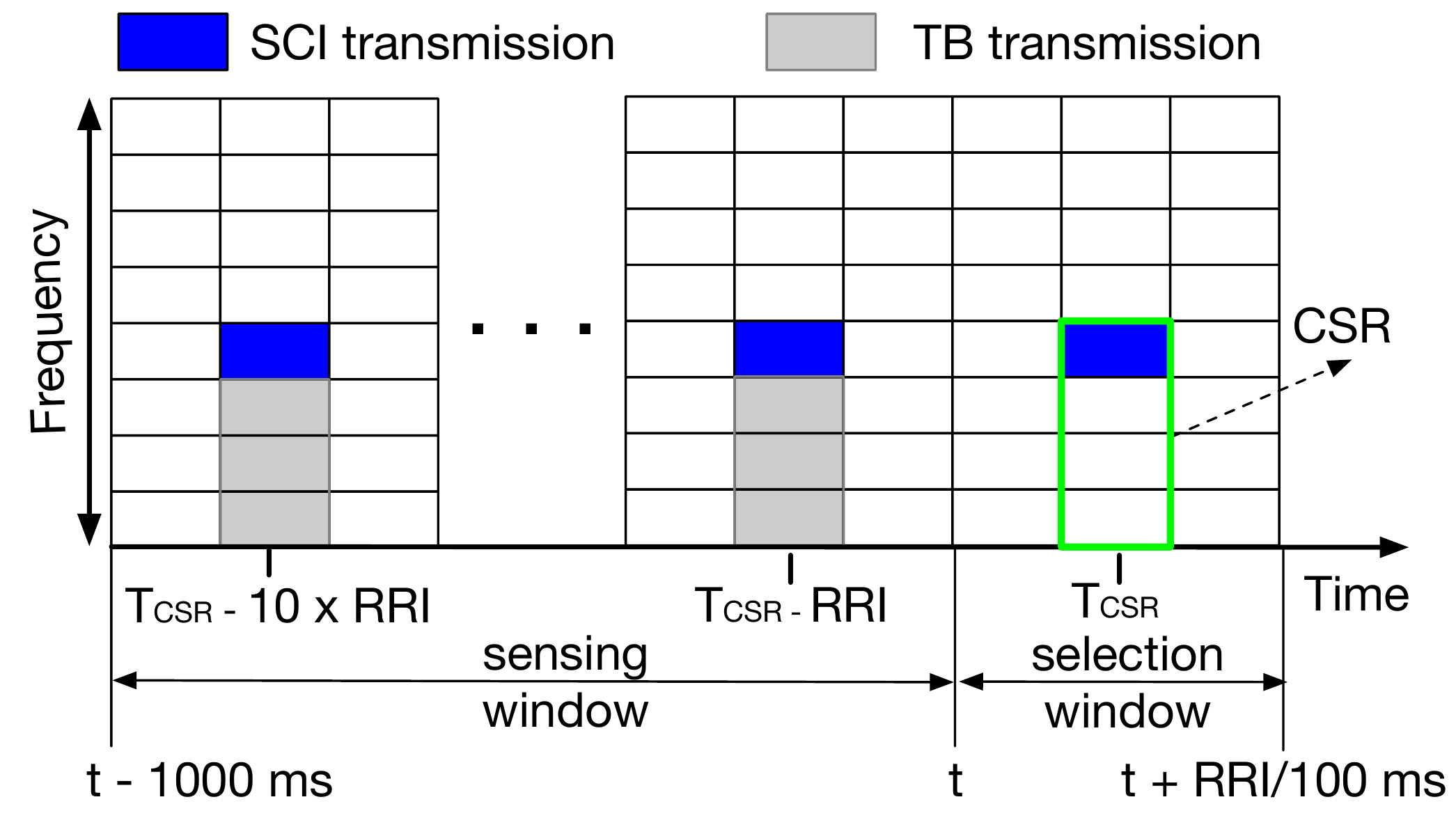}
\caption{CSR selection performed by the SB-SPS mechanism in C-V2X.}
\label{fig-macLayer}
\vspace{-15pt}
\end{figure}

The information contained in the grant is then passed to the PHY layer, where a list of subchannels known as Candidate Subframe Resources (CSRs) is created based on the grant specifications. This list contains all possible CSRs within a selection window comprising the period of time between the arrival of a packet from the application layer $t$ and the RRI or 100 ms (whichever is lower, in case the RRI is less than 100~ms). This is depicted in Figure \ref{fig-macLayer}. 

From this list, CSRs are discarded based on the information received during a sensing window of 1~s that spans between $t - 1000 ms$ and $t$ as shown in Figure \ref{fig-macLayer}. The PHY layer discards all CSRs whose SCIs indicate they will be reserved during the selection window as well as CSRs whose latest Reference Signal Received Power (RSRP) measurement is above a predefined threshold. The latter process is repeated by increasing the threshold by 3~dB until at least 20\% of all the CSRs in the selection window are available. After this process is finished, only 20\% of the remaining CSRs with the lowest Received Signal Strength Indicator (RSSI) are returned to the MAC layer, where a single CSR is selected at random.

Finally, at the MAC layer the SB-SPS mechanism signals the PHY layer to use the selected CSR to transmit the packet and decreases the RRC by one after each transmission. With this mechanism all upcoming packets are transmitted using the same CSR until the RRC reaches zero. When this occurs, the MAC layer can maintain the same reservation with a predefined probability of $P$ that can be chosen between $[0,0.8]$ or restart the entire process and generate a new grant.

\subsection{Default MCS selection in C-V2X}
\label{subsection-mcsSelection}

The specification for the access layer of C-V2X Release 14 states that the control messaging (SCI) must be transmitted with the lowest MCS (0), while data (TBs) can be transmitted using a minimum MCS of 0 or 3 and a maximum of 11. This assumes the vehicle travels at a speed less than 160 km/h \cite{etsi-accessLayer}. The specification does not mandate the use of a specific MCS within this range. However, it encourages any C-V2X evaluations to implement an MCS of QPSK with a coding rate of 0.5 as a baseline, or alternatively an MCS of QPSK with a coding rate of 0.7 or 16QAM with a coding rate of 0.5 \cite{3gpp-TS-36-885}.

Therefore, most studies in the area assume these values of MCS in order to evaluate the performance of C-V2X. With these assumptions, the MCS can be used in combination with the packet size to find the number of RBs $(N_PRB{})$, the TB size index $(I_{TBS})$ and the MCS index $(I_{MCS})$ using tables 7.1.7.2.1-1 and 8.6.1-1, provided in \cite{TB-sizes}. 

However, if the MCS is not assumed to be fixed for all TB transmissions and can be adapted in real-time, there are other important considerations that need to be taken into account, as now described.

\section{Modeling MCS adaptation in C-V2X}
\label{section-modeling}

Release 14 of the C-V2X standard does not preclude the adaptation of the MCS in the PSSCH within the specified ranges during ongoing transmissions and indeed suggests it as a potential congestion control mechanism \cite{3gpp-TR-21-916}. However, adapting the MCS during ongoing transmissions is not a trivial task and has several implications for the operation of the MAC and PHY layers of C-V2X that need to be considered in order to model MCS adaptation accurately.

\subsection{Impact of MCS selection on resource organization}
\label{subsection-channelConstraints}

The selection of different MCS configurations modifies the RB occupation and in some cases can modify the subchannel occupation in C-V2X due to its channelization structure.

According to the standard, the size of each subchannel and the number of RBs per subchannel depend on the available bandwidth and the PSCCH+PSSCH transmission scheme. In the case of an adjacent scheme, which is used by the standard to evaluate the performance of C-V2X in \cite{3gpp-TS-36-885}, the size of each subchannel in the PSSCH can be of 5, 6, 10, 15, 20, 25, 50, 75 or 100 RBs and the channel can contain 1, 3, 5, 8, 10, 15 or 20 subchannels. Despite the number of available configurations, ETSI indicates that the PSSCH should be configured with 5 subchannels of 10 RBs each for a channel bandwidth of 10 MHz \cite{etsi-accessLayer}. 

\begin{figure}[htbp]
\centering
\includegraphics[scale=0.40]{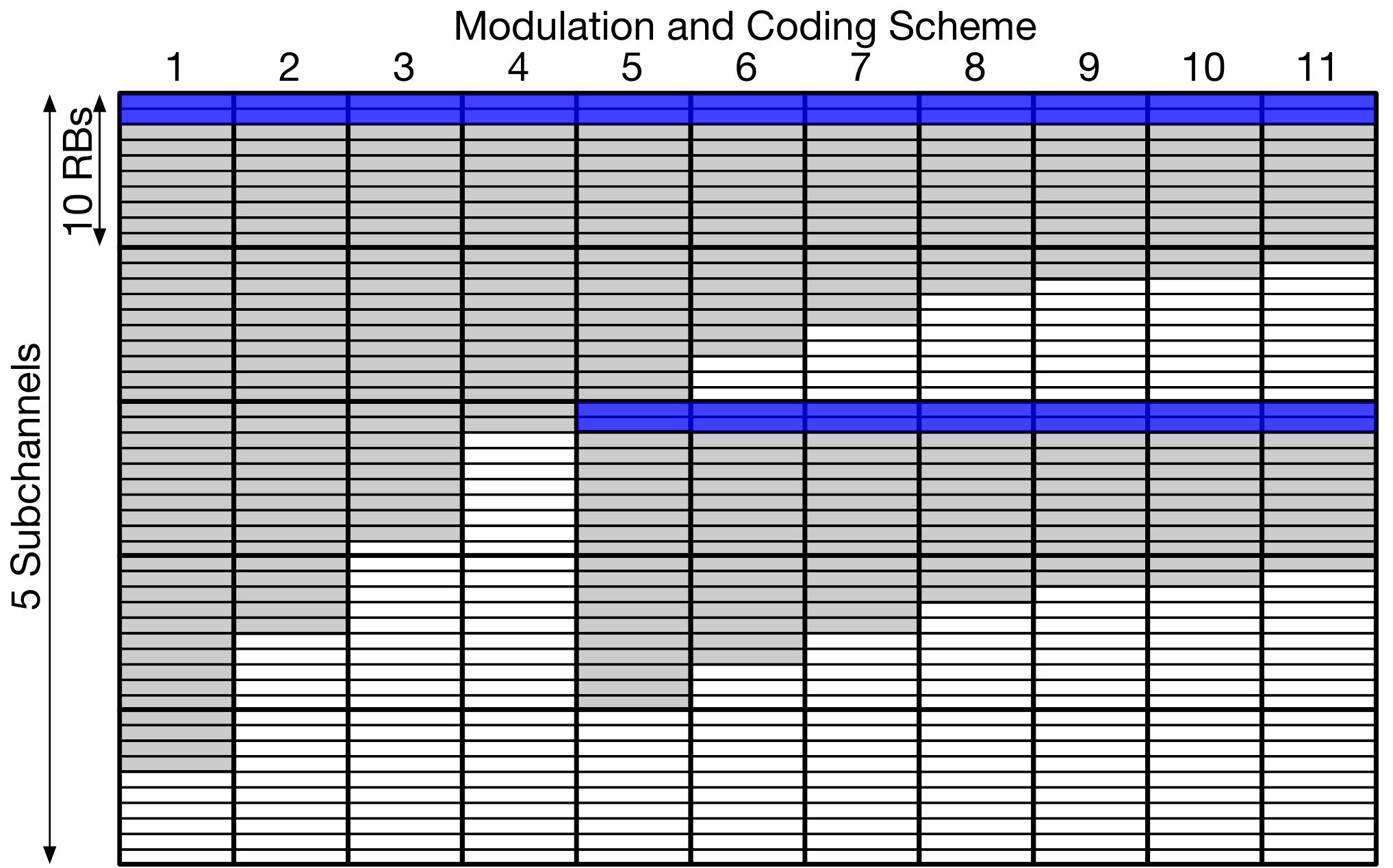}
\caption{Subchannel and RB occupation as a function of the MCS for a 190-Byte packet and 5 subchannels of 10 RBs.}
\label{fig-occupation190B}
\end{figure}

Under this configuration, the selection of MCS has a limited impact on the number of subchannels required to transmit a TB. This is depicted in Figure \ref{fig-occupation190B}, where the number of subchannels and RBs (including the 2 RBs of the SCI) required to transmit a 190-Byte packet has been obtained by following the procedure described in \cite{3gpp-TS-36-786}.

Figure \ref{fig-occupation190B} shows that the channelization structure of C-V2X limits the number of subchannels occupied when transmitting a 190-Byte packet to 5 (MCS 1), 4 (MCS 2), 3 (MCS 3 and MCS 4) or 2 (MCS 5 to MCS11 inclusive). However, Figure \ref{fig-occupation190B} also shows that the RB occupation varies for most MCS configurations independently of the subchannel occupation. This can be clearly seen for MCS 5 and MCS 11, which require the same number of subchannels (two) but occupy 20 and 11 RBs, respectively. These variations of RB occupancy as a function of the selected MCS and their implications need to be considered when MCS adaptation is implemented in practice.
 
\subsection{Impact of MCS adaptation on computation of power levels}
\label{subsection-powerConstraints}

As discussed, even if the number of required subchannels remains unchanged, the selection of MCS modifies the number of RBs required for transmission in most cases. This has implications on the computation of power levels in the channel and consequently on the operation of the SB-SPS mechanism in C-V2X that are important to consider.

For instance, if the total transmission power is assumed to be fixed, any variation in the number of used RBs directly modifies the transmission bandwidth and therefore, the Power Spectral Density (PSD) of each transmission. This needs to be considered every time the MCS is adapted to accurately compute the power levels in the channel and ensure that the PSD remains within the limit of 23 dBm/MHz as specified by the standard \cite{3gpp-TS-36-786}. This is often overlooked when analyzing the performance of C-V2X \cite{nxp-psd}.

More specifically, the variations of PSD due to MCS adaptation have an impact on the computation of important parameters such as the RSRP, RSSI and the Signal To Interference Ratio (SINR). RSRP is computed by obtaining the average power contribution of each Resource Element (RE) within the transmitted RBs \cite{3gpp-TS-36-214}. Therefore, if the number of RBs changes due to MCS adaptation, the value of RSRP must be computed accordingly.

The variations of PSD also have an impact on the levels of interference in the channel. For instance, a transmission that occupies a lower number RBs in a subchannel produces less interference than a transmission occupying the same subchannel but utilizing a higher number of RBs. This variation in interference has an impact on RSSI, which is computed as the received power plus the interference and thermal noise in the received SC-FDMA symbol \cite{3gpp-TS-36-214} and, therefore, it is directly related to the interference levels. This is also the case for the SINR, which is computed as the ratio between the received power and the interference plus the thermal noise. 

Based on this analysis, the accurate estimation of the received power, interference and thermal noise must be carefully considered when MCS adaptation is implemented. This is particularly important for parameters such as the RSRP and RSSI, which are used to perform CSR selection as well as for the SINR, which is used for decoding transmissions.

\subsection{Impact of MCS adaptation on CSR selection}
\label{subsection-csrSelection}

The C-V2X SB-SPS mechanism filters out CSRs based on the average RSRP and RSSI of the subchannels in the sensing window. Therefore, several scenarios can arise due to variations in RSRP and RSSI as a consequence of the implementation of MCS adaptation, with further implications for the selection process of CSRs.

\begin{figure}[htbp]
\centering
\includegraphics[scale=0.75]{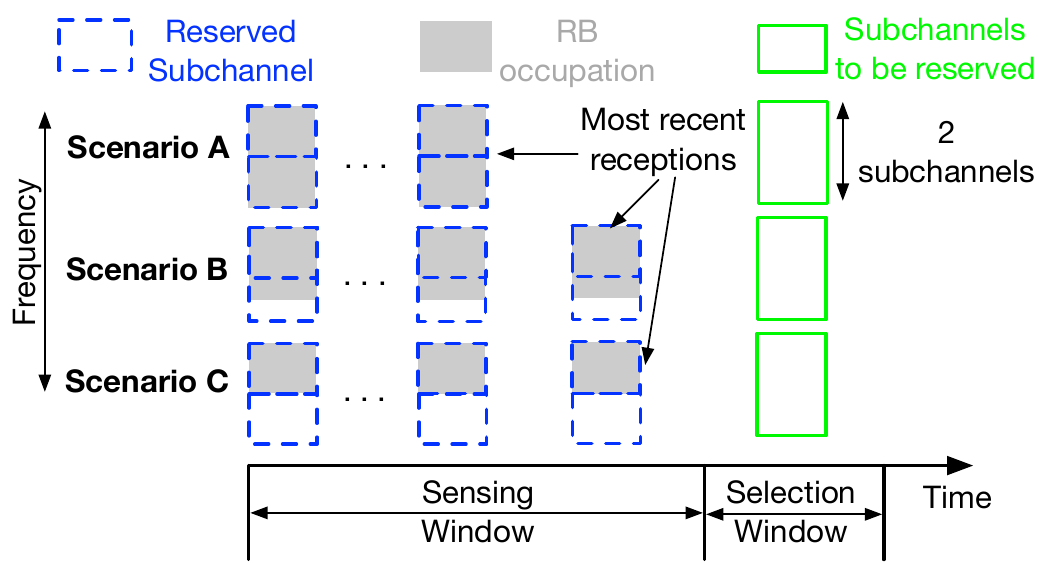}
\caption{Subchannel and RB occupation during CSR selection with MCS adaptation.}
\label{fig-csrFiltering}
\vspace{-15pt}
\end{figure}

Figure \ref{fig-csrFiltering} illustrates three potential scenarios depending on different reservation and occupation states of subchannels within the sensing window. Scenario A illustrates the case where two subchannels are reserved and fully occupied by a transmission where the utilized MCS requires use of all the RBs. As all the RBs in both subchannels are used, they both have the same levels of RSRP. This is also the case for the RSSI, because the occupied RBs contribute the same amount of interference to both subchannels.

However, in the case of Scenarios B and C, the levels of RSRP and RSSI of each subchannel in the sensing window can be different if the MCS adapts, thus varying the number of occupied RBs. For instance, the levels of RSRP in the upper subchannel of Scenario B, where all RBs are occupied, are higher than the lower subchannel. Thus, the upper subchannel has higher levels of RSSI, since there is a higher contribution of interference due to the use of a higher number of RBs. This also occurs in Scenario C, where two subchannels have been reserved but only the upper subchannel is fully occupied. In this scenario the lower channel has effectively zero RSRP since all the transmission power is used in the upper subchannel and also has lower RSSI because there is no interference.

Importantly, these scenarios also have implications during the selection of CSRs, because the SB-SPS mechanism filters subchannels based on RSRP and RSSI levels. SB-SPS first discards the subchannels with the highest RSRP measured during the most recent receptions within the sensing window, which is impacted by MCS adaptation as described for Scenarios B and C. After this process, SB-SPS selects only 20\% of the remaining subchannels with the lowest average RSSI computed. In this case, the average subchannel RSSI depends on the individual RSSI measurements in each subchannel, which can also vary as a function of the selected MCS as shown in scenarios B and C. 

Based on this analysis, it is evident that the variations in the subchannel RSRP and RSSI due to MCS adaptation have important implications in the selection of CSRs performed by the SB-SPS mechanism. Therefore, these implications have been considered in this study to accurately model the operation of MCS adaptation in C-V2X. 

\section{Performance evaluation of MCS adaptation for Congestion Control}
\label{section-performance}

MCS adaptation has been proposed as a mechanism for addressing distributed congestion control as proposed by 3GPP and ETSI \cite{etsi-congestionControl} and in the literature \cite{8080373,8795500}. We now carefully model and evaluate how MCS adaptation would perform if used for this purpose. 

The goal of distributed congestion control is to allow each vehicle to adjust its transmission parameters in order to reduce channel congestion. To perform congestion control each vehicle measures two metrics, the Channel Busy Ratio (CBR) and the Channel Occupancy Ratio (CR). The CBR measures the ratio of subchannels whose RSSI exceeded a predefined threshold during the last 100 ms and provides an estimation of the overall level of congestion in the channel. In contrast, the CR estimates the channel usage of each vehicle by measuring the ratio between the subchannels that have been used and reserved by the vehicle and all the available subchannels within a predefined period. Based on these two metrics, each vehicle can adjust its transmission parameters to reduce congestion if the CR exceeds a limit defined by the measured CBR. Recommended parameter values as defined by the standard \cite{etsi-accessLayer} are shown in Table \ref{table-cbrAndCrEtsi}.

\begin{table}[htbp]
\caption{CR limits per CBR range in C-V2X.}
\begin{center}
\begin{tabular}{c|c}
\hline
\textbf{Measured CBR} & \textbf{CR limit}\\
\hline
$CBR <= 0.3$ & no limit\\
$0.3 < CBR <= 0.65$ & 0.03\\
$0.65 < CBR <= 0.8$ & 0.06\\
$0.8 < CBR <= 0.1$ & 0.003\\
\hline
\end{tabular}
\label{table-cbrAndCrEtsi}
\end{center}
\end{table} 
\vspace{-12pt}

Using the CBR and CR, each vehicle can implement different mechanisms to reduce congestion, such as increasing the MCS. As described in Subsection \ref{subsection-channelConstraints}, increasing the MCS can reduce the number of required subchannels for a given transmission and therefore, reduce the CR. Moreover, even if the number of subchannels remains the same due to the channelization scheme of C-V2X, increasing the MCS reduces the number of required RBs which can reduce the average subchannel RSSI and ultimately the CBR.

We thus evaluate using MCS adaptation as a distributed congestion control mechanism, where each vehicle adapts the MCS depending on the measured CR and CBR. To this purpose, there are two MCS configurations available. MCS 7 is used by all vehicles by default when congestion is not detected in the channel (assuming a fixed packet size of 190B) and MCS 11 is implemented in the case where congestion conditions are met. With this mechanism, each vehicle increases the MCS from 7 to 11 when the CR exceeds the limit for its corresponding CBR and return to MCS 7 otherwise.
\vspace{-6pt}
\subsection{Simulation environment}
\label{subsection-setup}
\vspace{-5pt}
The performance of MCS adaptation for distributed congestion control was evaluated using OpenCV2X \cite{openCV2X,mccarthy2021opencv2x}, with parameters specified in \cite{3gpp-TS-36-885} and described in Table \ref{table-simConfiguration}. Three different vehicular densities are considered along with two fixed MCS configurations (MCS 7 and MCS 11) and an adaptive configuration, where the MCS alternates between both configurations depending on the congestion in the channel. Moreover, in order to maintain the PSD constant and within the limit of 23 dBm/MHz, the transmission power was also adapted depending on the selected MCS configuration. Specifically, the transmission power was set to 27.32 dBm for MCS 7 and 25.97 dBm for MCS 11. Finally, the application layer was configured for the transmission of 190 Bytes packets with a transmission frequency of 10 Hz. Results were obtained by averaging the outcome of 5 simulation runs.

\begin{table}[htbp]
\caption{Simulation scenario and parameters.}
\begin{center}
\begin{tabular}{c|c}
\hline
\textbf{Parameter} & \textbf{Value}\\
\hline
\multicolumn{2}{c}{\textbf{Vehicular topology}}\\
\hline
Road length & 2 km\\
Number of lanes & 6 (3 in each direction)\\
Lane width & 4 m\\
Vehicular density (veh/m) & 0.06, 0.09 and 0.20\\
\hline
\multicolumn{2}{c}{\textbf{Channel configuration}}\\
\hline
Carrier frequency & 5.9 GHz\\
Channel bandwidth & 10 MHz\\
Number of subchannels & 5 \\
Subchannel size & 10 RBs\\
\hline
\multicolumn{2}{c}{\textbf{Access \& Physical layers}}\\
\hline
Resource keep probability & 0\\
RSRP \& RSSI thresholds & -126 dBm \& -90 dBm\\
Propagation model & Winner+ B1\\
Noise figure & 9 dB\\
Shadowing variance LOS & 3 dB\\
MCS & 7, 11 and adaptive\\
Modulation and Coding Rate & QPSK/0.7, 16QAM/0.5 and adaptive\\
Transmission power & 27.32 dBm, 25.97 dBm and adaptive\\
\hline
\end{tabular}
\label{table-simConfiguration}
\vspace{-17pt}
\end{center}
\end{table}

\subsection{Results \& Analysis}
\label{subsection-results}

\begin{figure}[htbp]
\centering
\includegraphics[scale=0.42]{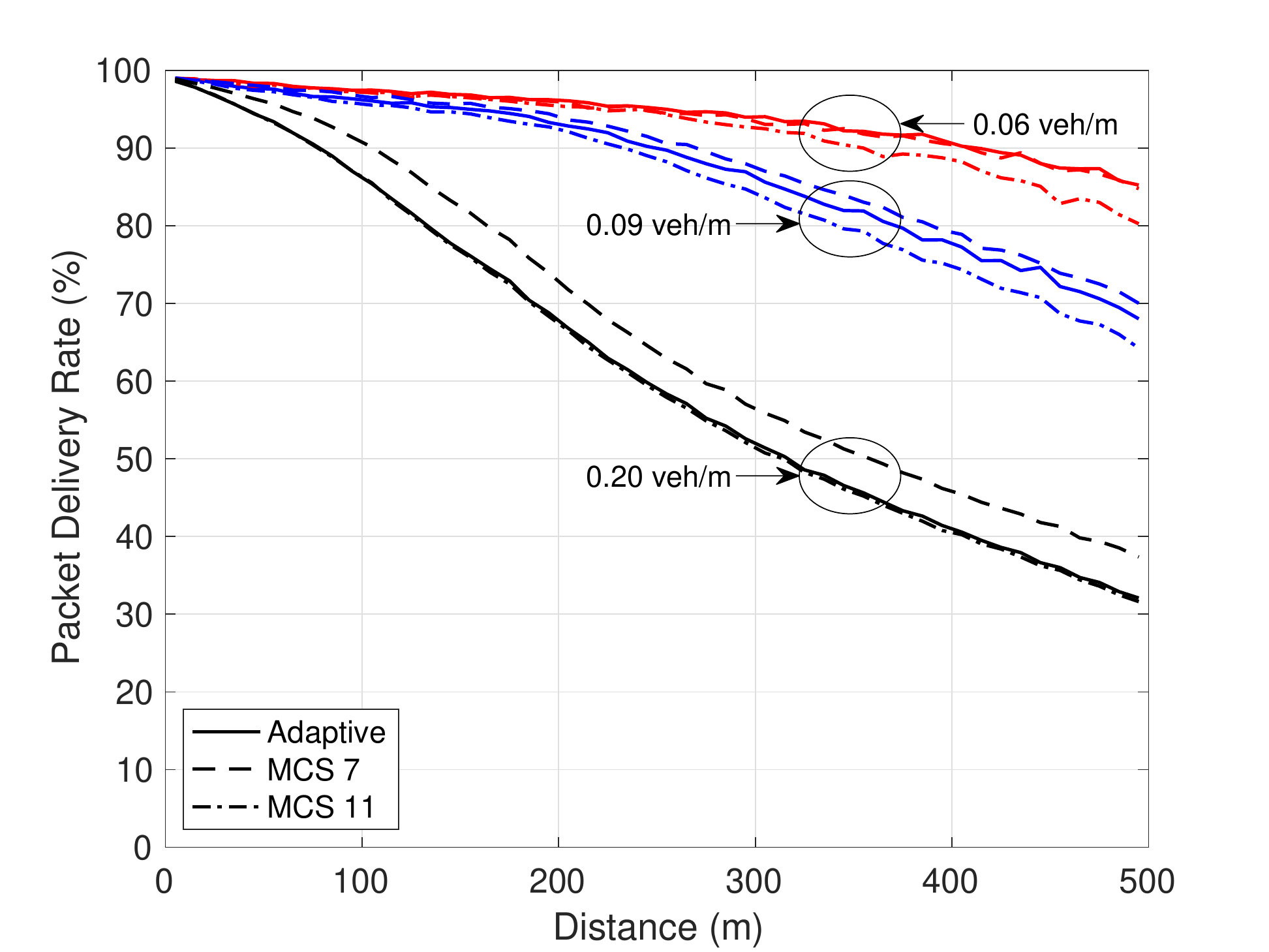}
\caption{Packet Delivery Rate.}
\label{fig-pdr}
\end{figure}

Figure \ref{fig-pdr} shows the Packet Delivery Rate (PDR) versus the distance for all MCS configurations and vehicular densities. As expected, the PDR decreases for MCS 7 and MCS 11 as the distance and vehicular density increase, due to the reduction in SINR and the increase of interference, respectively. It can also be noted that MCS 7 always out-performs MCS 11 with the difference in PDR increasing at higher vehicular densities. This can be attributed to the more robust modulation and coding rate of MCS 7 (QPSK/0.7), which out-performs MCS 11 (16QAM/0.5) at higher interference. 

\vspace{-15pt}
\vspace{-10pt}

Figure \ref{fig-pdr} also shows how the performance of MCS adaptation varies depending on the vehicular density. For instance, at 0.06 veh/m MCS adaptation achieves the same PDR as MCS~7 because the vehicular density is not high enough to trigger the congestion control mechanism. However, as the density increases to 0.09 veh/m, some vehicles in the scenario trigger MCS adaptation changing the MCS from 7 to 11, which slightly reduces the overall PDR. This is more evident at 0.20 veh/m, where MCS adaptation has the same performance as MCS 11 due to all vehicles increasing their MCS due to the higher vehicular density. Ultimately, it can be seen that adapting the MCS does nothing to improve PDR in response to congestion. This is due to no alteration in the sub-channel occupation although RB occupation is reduced. 

\begin{figure}[htbp]
\centering
\includegraphics[scale=0.42]{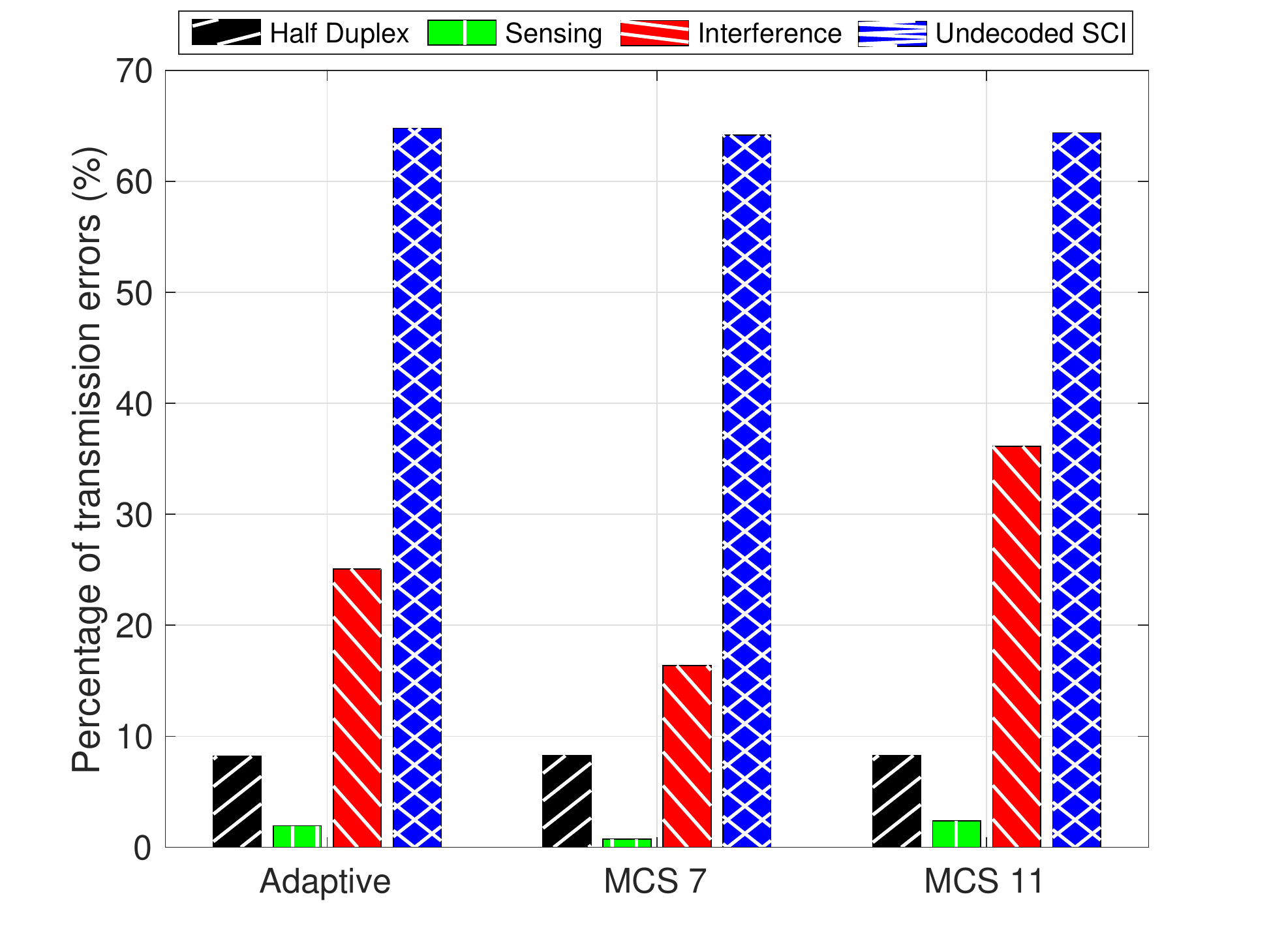}
\caption{Percentage of transmission errors versus cause of error for all MCS configurations at 0.09 veh/m.}
\label{fig-TbErrors}
\end{figure}

Figure \ref{fig-TbErrors} shows the percentage of transmission errors versus the cause of error for all MCS configurations. The results are shown for a density of 0.09 veh/m and correspond only to errors within a range of 500~m. It can be seen that the percentage of half duplex errors is the same for all MCS configurations. This is expected as the transmission rate for MCS 7 and MCS 11 is the same, which results in the same number of lost transmissions due to half duplex limitations. In terms of sensing errors, MCS 7 performs slightly better than MCS 11. This occurs because MCS 7 can implement a higher transmission power than MCS 11 while maintaining the PSD within the 23 dBm/MHz limit. Specifically, the transmission power can be set to 27.32 dBm for MCS 7 and 25.97 dBm for MCS 11 because MCS 7 occupies a larger bandwidth (15 RBs) than MCS 11 (11 RBs). In the case of MCS adaptation, the percentage of sensing errors is within the range of MCS 7 and MCS 11 as some vehicles increase their MCS when the mechanism is triggered at the configured density.

\vspace{-5pt}
\vspace{-8pt}

In terms of transmission errors due to interference, the more robust MCS 7 (QPSK/0.7) out-performs MCS 11 (16QAM/0.5) when the same levels of interference are present in the channel. In the case of MCS adaptation, the errors due to interference are within the range of MCS 7 and MCS 11 as the MCS is increased in some of the vehicles in the scenario.

Finally, Figure \ref{fig-TbErrors} shows the percentage of transmission errors due to undecoded SCIs. This type of error occurs when the SCI cannot be decoded at the receiver, resulting in the corresponding TB being immediately discarded. This is attributable to the fact that errors are analyzed within a 500 m range, where interference from SCI collisions is more likely to cause undecoded SCIs and impact TB transmissions. Figure \ref{fig-TbErrors} also shows that the selection of different MCS configurations has no impact on SCI collisions, as the MCS only modifies the number of RBs for the TB and not the SCI. 

\begin{figure}[htbp]
\centering
\includegraphics[scale=0.42]{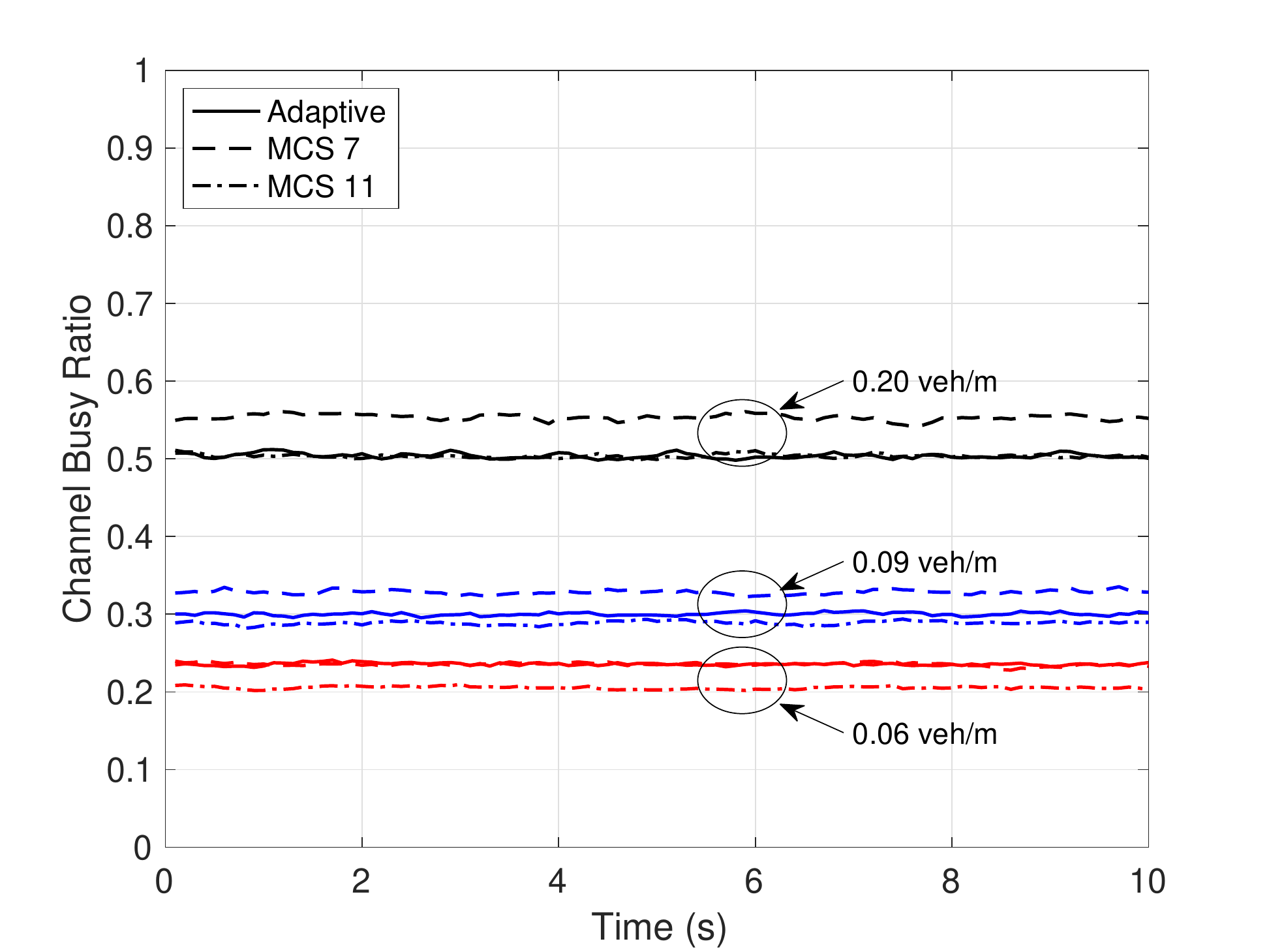}
\caption{Channel Busy Ratio.}
\label{fig-cbr}
\vspace{-15pt}
\end{figure}

\begin{figure}[htbp]
\centering
\includegraphics[scale=0.42]{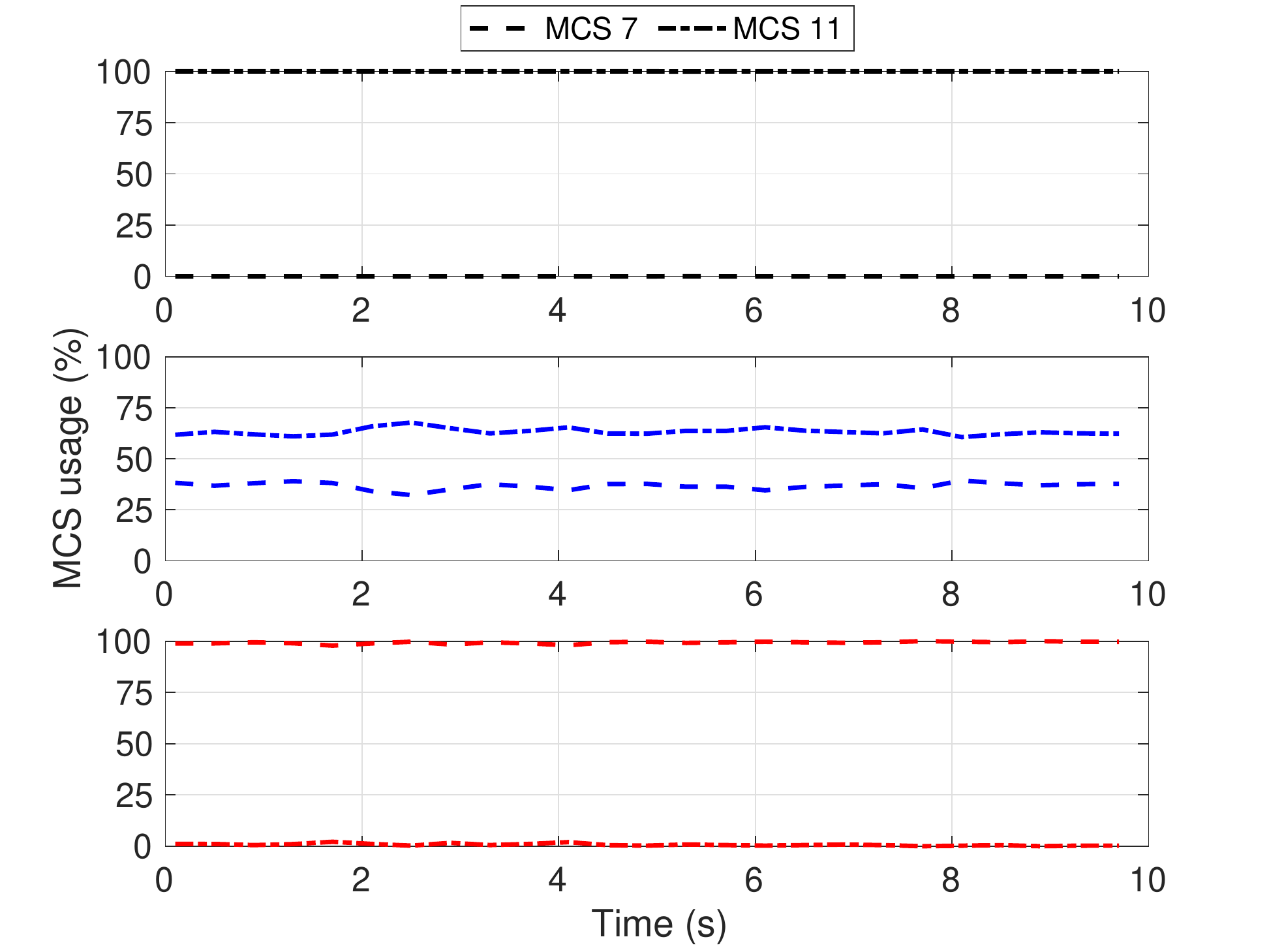}
\caption{MCS usage for MCS adaptation at 0.06 veh/m (lower figure), 0.09 veh/m (middle figure) and 0.20 veh/m (upper figure).}
\label{fig-mcsUsage}
\vspace{-8pt}
\end{figure}

To validate the operation of MCS adaptation for congestion control, Figure \ref{fig-cbr} shows the CBR for all vehicular densities and MCS configurations while Figure \ref{fig-mcsUsage} shows the MCS usage for MCS adaptation for all vehicular densities. 

\vspace{-14pt}

In Figure \ref{fig-cbr}, the CBR levels at 0.06 veh/m are lower than 0.3, which is the minimum CBR required to trigger MCS adaptation according to Table~\ref{table-cbrAndCrEtsi}. Therefore, at 0.06 veh/m and with MCS adaptation enabled almost all vehicles in the scenario implement MCS 7, as shown in Figure \ref{fig-mcsUsage}. It can also be seen that for 0.09 veh/m, the CBR increases around the minimum CBR limit of 0.3 which is enough to trigger MCS adaptation. This is validated in Figure \ref{fig-mcsUsage} where it is shown that around 60\% of vehicles increase their MCS to 11, while the remaining 40\% implement MCS 7. At 0.20 veh/m, Figure \ref{fig-cbr} also shows that the CBR reaches 0.5 as its highest value, which is higher than the minimum CBR limit. Consequently, MCS adaptation is triggered in all vehicles of the scenario, which increase their MCS to 11 as shown in Figure \ref{fig-mcsUsage}. 

Importantly, the results of Figure \ref{fig-cbr} indicate that the CBR levels are generally lower for MCS 11 across all vehicular densities. This is the result of the reduction in subchannel RSSI that occurs due to the lower RB occupation of the higher MCS'. Moreover, this result highlights the main advantage of increasing the MCS in congested scenarios, which is an overall reduction of channel congestion.

Finally, in the case of MCS adaptation, Figure \ref{fig-cbr} shows how the CBR levels vary depending on the vehicular density. For instance, at 0.06 veh/m the CBR of MCS 7 and MCS adaptation are practically the same due to the lower density in the scenario, which is below the CBR limit required to trigger MCS adaptation. As the density increases to 0.09 veh/m, MCS adaptation is able to reduce the overall CBR, as 60\% of the vehicles in the scenario switch to MCS 11 to cope with the higher congestion in the channel. This trend is even clearer at 0.20 veh/m, where all vehicles switch to MCS 11 due to the high congestion, achieving a higher reduction on CBR.

\begin{figure}[htbp]
\centering
\includegraphics[scale=0.42]{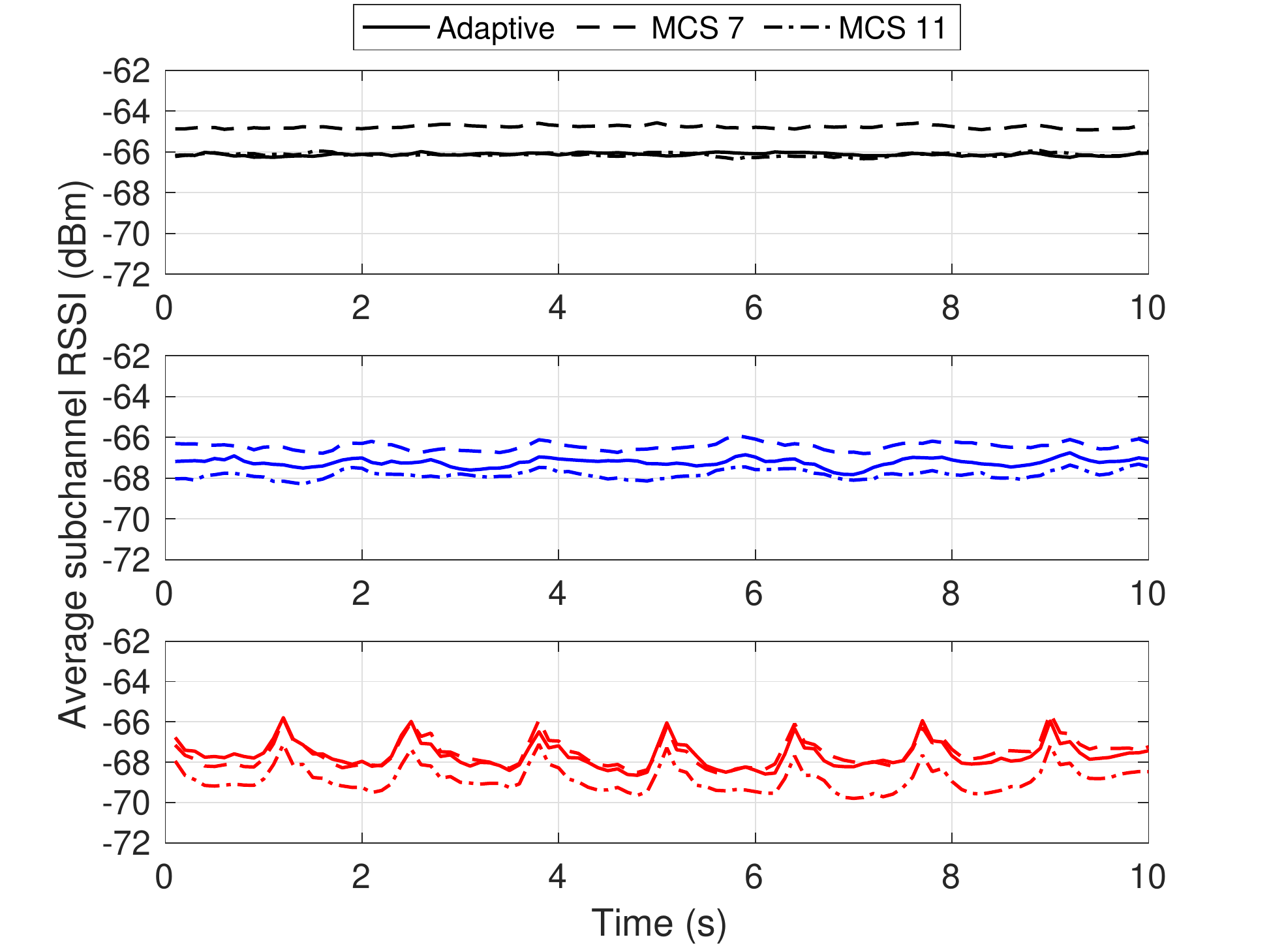}
\caption{Average subchannel RSSI at 0.06 veh/m (lower figure), 0.09 veh/m (middle figure) and 0.20 veh/m (upper figure).}
\label{fig-averageRssi}
\end{figure}

This reduction in CBR can be validated through Figure \ref{fig-averageRssi} where the average subchannel RSSI is shown for all MCS configurations and vehicular densities. It can be seen that the average subchannel RSSI increases with the vehicular density due to the higher levels of interference in the scenario. More importantly, it shows that the average subchannel RSSI of MCS 7 is higher than MCS 11 for all the vehicular densities, indicating that the higher RB occupation of MCS 7 with respect to MCS 11 increases the subchannel RSSI and ultimately the CBR, as shown in Figure \ref{fig-cbr}.

\vspace{-8pt}
\vspace{-8pt}

Figure \ref{fig-averageRssi} also shows how the average subchannel RSSI of MCS adaptation depends on the configured vehicular density. For instance, at 0.06 veh/m the number of vehicles in the scenario is not enough to trigger MCS adaptation, and therefore subchannel RSSI is the same as MCS 7. However, as the density increases to 0.09 veh/m and 0.20 veh/m the average subchannel RSSI is lower for MCS adaptation in comparison with MCS 7 because the vehicles in the scenario increase their MCS due to the higher congestion in the channel.

\section{Conclusions}
\label{section-conclusions}

This paper comprehensively studies the implications of implementing MCS adaptation in the C-V2X sidelink. Specifically, it analyses the impact on the resource configuration, power levels and in the operation of the SB-SPS mechanism. To validate the study, MCS adaptation is evaluated as a mechanism for congestion control as proposed by ETSI. The results indicate that under high vehicular density MCS adaptation is capable of reducing the subchannel interference and the  channel congestion. However, the results also highlight the limitations due to the channelization constraints of C-V2X and demonstrate that the technique does not improve the overall performance of C-V2X under the analyzed congestion conditions. Despite the limitations of MCS adaptation for congestion control, the study provides valuable insights on the operation of the technique in C-V2X, laying the foundation for other applications that can benefit from its application, such as the transmission of packets of variable size.
\vspace{-3pt}
\bibliographystyle{./bibliography/IEEEtran.bst}
\bibliography{./bibliography/references}

\begin{thebibliography}{10}
\providecommand{\url}[1]{#1}
\csname url@samestyle\endcsname
\providecommand{\newblock}{\relax}
\providecommand{\bibinfo}[2]{#2}
\providecommand{\BIBentrySTDinterwordspacing}{\spaceskip=0pt\relax}
\providecommand{\BIBentryALTinterwordstretchfactor}{4}
\providecommand{\BIBentryALTinterwordspacing}{\spaceskip=\fontdimen2\font plus
\BIBentryALTinterwordstretchfactor\fontdimen3\font minus
  \fontdimen4\font\relax}
\providecommand{\BIBforeignlanguage}[2]{{%
\expandafter\ifx\csname l@#1\endcsname\relax
\typeout{** WARNING: IEEEtran.bst: No hyphenation pattern has been}%
\typeout{** loaded for the language `#1'. Using the pattern for}%
\typeout{** the default language instead.}%
\else
\language=\csname l@#1\endcsname
\fi
#2}}
\providecommand{\BIBdecl}{\relax}
\BIBdecl

\bibitem{4526014}
D.~Jiang and L.~Delgrossi, ``Ieee 802.11p: Towards an international standard
  for wireless access in vehicular environments,'' in \emph{VTC Spring 2008 -
  IEEE Vehicular Technology Conference}, 2008, pp. 2036--2040.

\bibitem{3gpp-TS-36-885}
3GPP, \emph{Study on LTE-based V2X services (v14.0.0)}, Jul. 2016, tR 38.885.

\bibitem{3gpp-TR-21-915}
3GPP, \emph{Release 15 Description (v15.0.0)}, Sep. 2019, tR 21.915.

\bibitem{3gpp-TR-21-916}
3GPP, \emph{Release 16 Description (v16.0.0)}, Jun. 2021, tR 21.916.

\bibitem{etsi-congestionControl}
ETSI, \emph{Congestion Control Mechanisms for the C-V2X PC5 interface; Access
  layer part (V1.1.1)}, Nov. 2018.

\bibitem{8676227}
A.~Bazzi, A.~Zanella, and B.~M. Masini, ``Optimizing the resource allocation of
  periodic messages with different sizes in lte-v2v,'' \emph{IEEE Access},
  vol.~7, pp. 43\,820--43\,830, 2019.

\bibitem{8080373}
R.~Molina-Masegosa and J.~Gozalvez, ``Lte-v for sidelink 5g v2x vehicular
  communications: A new 5g technology for short-range vehicle-to-everything
  communications,'' \emph{IEEE Vehicular Technology Magazine}, vol.~12, no.~4,
  pp. 30--39, 2017.

\bibitem{8795500}
A.~Mansouri, V.~Martinez, and J.~Härri, ``A first investigation of congestion
  control for lte-v2x mode 4,'' in \emph{2019 15th Annual Conference on
  Wireless On-demand Network Systems and Services}, 2019, pp. 56--63.

\bibitem{9048738}
A.~Bazzi, ``Congestion control mechanisms in ieee 802.11p and sidelink c-v2x,''
  in \emph{2019 53rd Asilomar Conference on Signals, Systems, and Computers},
  2019, pp. 1125--1130.

\bibitem{9133075}
R.~Molina-Masegosa, J.~Gozalvez, and M.~Sepulcre, ``Comparison of ieee 802.11p
  and lte-v2x: An evaluation with periodic and aperiodic messages of constant
  and variable size,'' \emph{IEEE Access}, vol.~8, pp. 121\,526--121\,548,
  2020.

\bibitem{6823631}
Y.~Yao, X.~Zhou, and K.~Zhang, ``Density-aware rate adaptation for vehicle
  safety communications in the highway environment,'' \emph{IEEE Communications
  Letters}, vol.~18, no.~7, pp. 1167--1170, 2014.

\bibitem{7480406}
Y.~Yao, X.~Chen, L.~Rao, X.~Liu, and X.~Zhou, ``Lora: Loss differentiation rate
  adaptation scheme for vehicle-to-vehicle safety communications,'' \emph{IEEE
  Transactions on Vehicular Technology}, vol.~66, no.~3, pp. 2499--2512, 2017.

\bibitem{5487432}
J.~Camp and E.~Knightly, ``Modulation rate adaptation in urban and vehicular
  environments: Cross-layer implementation and experimental evaluation,''
  \emph{IEEE/ACM Transactions on Networking}, vol.~18, no.~6, pp. 1949--1962,
  2010.

\bibitem{etsi-accessLayer}
ETSI, \emph{Intelligent Transport Systems (ITS); Access layer specification for
  Intelligent Transport Systems using LTE Vehicle to everything communication
  in the 5,9 GHz frequency band (V1.1.1)}, Nov. 2018.

\bibitem{TB-sizes}
ETSI, \emph{LTE; Evolved Universal Terrestrial Radio Access (E-UTRA) Physical
  layer procedures (V14.17.0 Release 14)}, June. 2021.

\bibitem{3gpp-TS-36-786}
3GPP, \emph{User Equipment (UE) radio transmission and reception (v14.0.0,
  Release 14)}, Mar. 2017, tR 38.786.

\bibitem{nxp-psd}
NXP, \emph{Analysis of CAMP Report on the C-V2X Performance Assessment
  Project}, 2019.

\bibitem{3gpp-TS-36-214}
3GPP, \emph{Physical layer Measurements (v14.0.0)}, Dec. 2017, tS 36.214.

\bibitem{openCV2X}
B.~{McCarthy} and A.~{O'Driscoll}, ``Opencv2x mode 4: A simulation extension
  for cellular vehicular communication networks,'' in \emph{2019 IEEE 24th
  International Workshop on Computer Aided Modeling and Design of Communication
  Links and Networks (CAMAD)}, 2019, pp. 1--6.

\bibitem{mccarthy2021opencv2x}
\BIBentryALTinterwordspacing
B.~McCarthy, A.~Burbano-Abril, V.~R. Licea, and A.~O'Driscoll, ``Opencv2x:
  Modelling of the v2x cellular sidelink and performance evaluation for
  aperiodic traffic,'' \emph{CoRR}, vol. abs/2103.13212, March 2021. [Online].
  Available: \url{https://arxiv.org/abs/2103.13212}
\BIBentrySTDinterwordspacing

\end{thebibliography}


\end{document}